# ParaSail: A Pointer-Free Pervasively-Parallel Language for Irregular Computations

S. Tucker Taft[a]

a   AdaCore, Lexington, MA, USA

**Abstract**   ParaSail is a language specifically designed to simplify the construction of programs that make full, safe use of parallel hardware even while manipulating potentially irregular data structures. As parallel hardware has proliferated, there has been an urgent need for languages that ease the writing of correct parallel programs. ParaSail achieves these goals largely through simplification of the language, rather than by adding numerous rules. In particular, ParaSail eliminates global variables, parameter aliasing, and most significantly, re-assignable pointers. ParaSail has adopted a pointer-free approach to defining complex data structures. Rather than using pointers, ParaSail supports flexible data structuring using expandable (and shrinkable) objects implemented using region-based storage management, along with generalized indexing. By eliminating global variables, parameter aliasing, and pointers, ParaSail reduces the complexity for the programmer, while still allowing ParaSail to provide flexible, pervasive, safe, parallel programming for irregular computations. Perhaps the most interesting discovery in this language development effort, based on over six years of use by the author and a group of ParaSail users, has been that it is possible to simultaneously simplify the language, support parallel programming with advanced data structures, and maintain flexibility and efficiency.

## The Art, Science, and Engineering of Programming

**ParaSail: A Pointer-Free Pervasively-Parallel Language**

## 1 Introduction and History

ParaSail is a language specifically designed to simplify the construction of programs that make full, safe use of parallel hardware even while manipulating potentially irregular data structures. The default evaluation semantics for expressions are parallel, and even at the statement level parallel semantics are the default in the absence of data dependences, all of which are immediately visible to the compiler using only local analysis. Safety is achieved in ParaSail largely through simplification of the language, that is, by removing features that interfere with straightforward automatic and safe parallelization. In particular, functions have no access to global variables, and aliasing of any two function parameters is not permitted if either is updatable within the function. Furthermore, parameter passing uses a *hand-off* semantics analogous to that of the Hermes language [32], such that when a variable is passed to one function, it may not be manipulated further by the caller nor passed to any other function if the first function is given read/write access to the parameter; or if the first function is given read-only access, then it may not be passed to any other function with read/write access. Finally, there are no user-visible re-assignable pointers in the language, further simplifying aliasing analysis. The net effect is that in ParaSail, expressions such as F(X) + G(Y) can be safely evaluated in parallel, without the compiler looking inside the bodies of the functions F or G.

The design of ParaSail began in September 2009, and was documented along the way in a web log [34]. The first interpreter-based implementation of the language began in 2011, and was largely completed by 2012. In 2014, an LLVM-based [21] code generator was written (in ParaSail itself). At the same time, an integrated static analysis capability was developed (in ParaSail) to provide more advanced compile-time error messages, as well as identify for the compiler places where run-time checks should be inserted to ensure safe execution.

The ParaSail front end generates instructions for a virtual machine specifically designed for pervasively parallel semantics (ParaSail Virtual Machine – PSVM) – see figure 1. These PSVM instructions can be directly executed in the ParaSail interpreter, or can be translated by the ParaSail compiler to an LLVM representation and then to machine code. The PSVM instructions may also be analysed statically, using an analyser called ParaScope, based on a variant of abstract interpretation [11, 27]. Parsers for parallel variants of other languages have been built which share the ParaSail abstract syntax tree (AST), semantic analysis phase, and all later code-generation phases. There are parsers for variants of Java (Javallel), Python (Parython), and the SPARK [9] subset of Ada (Sparkel).

The remainder of this paper is organized as follows: Section 2 introduces the ParaSail language through a series of examples. Some of the examples include a fair amount of code, to illustrate how realistically-sized modules appear in ParaSail, and to show the kinds of operations that might be used to support irregular data structures such as hash tables. Section 3 describes the ParaSail pointer-free region-based storage model and its implementation. Section 4 describes the ParaSail model for safe parallel execution, and its implementation, and provides some examples of irregular computations. Section 5 describes related work. Section 6 evaluates ParaSail's features relative to other parallel programming languages, and summarizes the contributions of the ParaSail design.

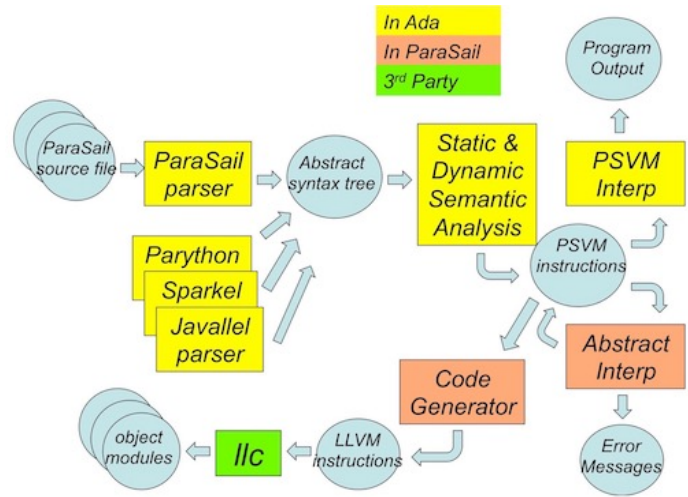

**Figure 1** ParaSail Tool Chain

## 2 The ParaSail Language

ParaSail is a relatively conventional looking object-oriented language, but with certain restrictions and features that make it particularly appropriate for automatic parallelization. ParaSail is also somewhat unusual in its heavy use of syntactic sugar [20] to integrate user-defined types into the special syntactic constructs of the language. That is, rather than providing special syntactic constructs, such as literals or indexing, to only a set of built-in types (such as the numeric types) or built-in type constructors (such as those for array types), ParaSail ties such special syntactic constructs to specific operators, which can be defined on any type. For example, uses of the indexing syntactic construct A[I] is expanded (*de-sugared*) into a call on the "indexing" operator as "indexing"(A, I), and any type with an "indexing" operator supports the A[I] syntax.

### 2.1 ParaSail Modules

A ParaSail program is composed of a set of hierarchically named *modules*, along with one or more stand-alone *operations*. A *module* defines the data components and the operations that together encapsulate a portion of the program's logical structure. An *operation* is a callable entity, referred to as a function or a procedure in many languages. A module consists of an **interface**, which defines the externally visible data components and operations of the module, and, optionally, a **class** that implements the interface and defines any internal data components and operations. A module whose interface part is declared as **abstract** (meaning it is a pure interface and has no implementation), or a module that has no operations needing implementation, need not have a class part. A module may be declared to extend another module, in which case it inherits both code (operations) and data (components) from that parent module. A module may also be declared to *implement* one or more other module's interfaces. Note that even if a module has its own class part, its interface may be implemented by other modules. In other words, ParaSail supports single inheritance of implementation (by extending a parent module) and multiple inheritance of interfaces (by implementing one or more other modules' interfaces).

The interface of a module declares its module parameters (types or values), defines zero or more visible data components, and declares zero or more visible operations. Each component is declared as either a **const** or a **var** component, determining whether the component must be defined once upon object creation or may be updated multiple times during the life of an enclosing object. The class part of a module defines (in an internal part) zero or more additional components, and zero or more additional (internal) operations. The internal part is followed by the word **exports** and then by the implementation (body) of each of the visible operations declared in the interface.

## 2.2 ParaSail Types and Objects

A ParaSail *type* is defined by instantiating a ParaSail module, which consists of specifying the name of the module and then an actual type or value for each formal parameter of the module. For example, Set<Integer> defines a set-of-integers type given a Set module with one type formal parameter, presuming Integer is itself a type. A named type may be declared by specifying the type name and the type definition; for example: **type** Int_Set **is** Set<Integer>. A ParaSail *object* is an instance of a type, and is declared as either a **var** (*variable*) or **const** (*constant*) object. For example:

```
1    var X: Set<Integer> := []
2    const Y: Set<Integer> := [35, 42]
3    var Z := Y
```

Note that the type need not be specified if the type of the initial value can be resolved unambiguously. An initial value need not be provided for a **var** object, but then a type specification is clearly required. When specifying the type for an object, the keyword **optional** may be specified. Each ParaSail type includes one additional value called **null**, but only objects (or components) specified as **optional** are permitted to be assigned a null value.

An object (or component) with a null value takes up minimal space. The space for the object grows automatically when the object is assigned a non-null value, and shrinks again when assigned back to a null value. See below for a further discussion of the ParaSail region-based storage model.

## 2.3 ParaSail Map Module Example

Here is an example of the interface for a Map module which provides a mapping from keys to values. Hashable is itself another (abstract) interface, which the Key_Type must implement. As described above, an interface begins with a list of formal parameters, in this case Key_Type and Value_Type, followed by definitions of any visible data components (none in this case) and visible operations (such as "[]", "indexing", and Is_Empty). In this case we also see the definition of a local type Pair, which is used in the definition of various operations that follow. Pair is defined as an instantiation of the module Key_Value, which also has two parameters. The full declaration of the interface to module Key_Value is given in section 2.3.4. As illustrated below, operations can either be declared with **op** (*operator*) to indicate they are invoked using special syntax (*syntactic sugar*), or with **func** (*function*) to indicate they are invoked using normal name(params) syntax.

```
1  interface PSL::Containers::Map // A hashed-map module
2    <Key_Type is Hashable<>; Value_Type is Assignable<>> is
3      type Pair is Key_Value<Key_Type, Value_Type>
4      op "[]"() -> Map
5          // Return an empty map
6      op "|="(var M: Map; KV: Pair)
7          // Add Key=>Value to Map
8      op "|"(M: Map; KV: Pair) -> Map
9          // Return Map with Key=>Value added
10     op "+="(var M: Map; KV: Pair) is "|="
11         // A synonym
12     op "in"(Key: Key_Type; M: Map) -> Boolean
13         // True if Key is in map
14     op "-="(var M: Map; Key: Key_Type)
15         // Remove mapping for Key, if any
16     op "index_set"(M: Map) -> Set<Key_Type>
17         // Return set of keys with mappings
18     op "indexing"(ref M: Map; Key: Key_Type) {Key in M} -> ref Value_Type
19         // Used for references to M[Key]; {} is precondition.
20     func Remove_Any(var M: Map) -> optional Pair
21         // Remove mapping from Map; Return null if map is empty.
22     op "magnitude"(M: Map) -> Univ_Integer
23         // Number of mappings in the table; Supports "|M|" notation
24     func Is_Empty(M: Map) -> Boolean
25         // True if no mappings in the table
26 end interface PSL::Containers::Map
```

### 2.3.1 Parameterized and Parameterless Modules

The list of module parameters is given in <> immediately after the module name. Each parameter that is a type specifies its name and the interface it must implement. Assignable imposes no special interface requirements. Hashable is defined as follows:

```
1  abstract interface PSL::Containers::Hashable<> is
2      op "=?"(Left, Right: Hashable) -> Ordering
3      func Hash(Val: Hashable) -> Univ_Integer
4  end interface PSL::Core::Hashable
```

Hashable has no module parameters (empty <>). The parameterless interfaces that a module implements need not be specified explicitly – any module that provides the required operations of a parameterless interface is automatically defined to implement the interface. This allows a parameterized module to define exactly the operations it needs of a given type parameter using a parameterless interface, without requiring that the actual type used to instantiate the module explicitly lists that interface as one of its implemented interfaces. Hence, if a hypothetical type Integer provides the "=?" and Hash operations, then Map<Integer, String> is permitted, even if the module defining Integer doesn't explicitly mention Hashable among the modules that it implements.

### 2.3.2 The "=?" Operator

The "=?" operator in Hashable is used to define the equality and relational properties of the associated type. Syntactic sugar is used to map both of the equality operators ("==" and "!=") and all four of the relational operators ("<", "<=", ">=", ">") to this one compare operator. The Ordering result type of "=?" is an enumeration type with four possible values, #less, #equal, #greater, and #unordered. The expression A == B is de-sugared into (A =? B) **in** [#equal] while A <= B is de-sugared into (A =? B) **in** [#less, #equal], where [X, Y, ...] is a *container aggregate* notation used to define values for sets, maps, arrays, etc. The container aggregate notation is itself syntactic sugar, which de-sugars into a sequence of operations starting with "[]" to produce an empty container of the desired type, and then invocations of operators to add elements sequentially to the growing container. These operations are performed at compile-time if all elements are compile-time-known (e.g. literals) or instantiation-time known (e.g. module parameters). Note that "=?" can be used for partially-ordered types as well as fully ordered types. If A =? B returns #unordered, then both A <= B and A >= B will return #false, indicating that A and B are not ordered with respect to one another. The uniform use of syntactic sugar for all comparisons as well as all aggregates means that user-defined types can be as powerful and usable as any built-in scalar or container type.

### 2.3.3 The Map Class

A module may have a class part that provides an implementation of its interface. Below is a partial listing of the Map class. As described above, a class starts with declarations of internal data components (Table in this case) and internal operations (none needed in this case), followed by the keyword **exports** and then the implementations of the visible operations of the module.

```
1  class PSL::Containers::Map is
2      var Table: Hash_Table<Pair> // A Map is a wrapper of a Hash_Table of Key=>Value pairs
3    exports
4      op "[]"() -> Map is
5          // Return an empty map
6          return (Table => [])
7      end op "[]"
8      op "|="(var Left: Map; Right: Pair) is
9          // Add Key=>Value to Map
10         Left.Table |= Right
11     end op "|="
12     op "|"(Left: Map; Right: Pair) -> Result: Map is
13         // Return Map with Key=>Value added
14         Result := Left
15         Result |= Right
16     end op "|"
17     ...
18 end class PSL::Containers::Map
```

The above is an example of a wrapper module, meaning it has exactly one component, while providing a different set of operations than the underlying data object. ParaSail is designed so that no per-object space overhead is necessary for such a wrapper module – it can have exactly the same run-time representation as its underlying component, even though its compile-time interface is potentially quite different. This contrasts with some languages where each layer of compile-time abstraction becomes another layer of run-time object overhead [26].

The implementation of the "|" operator above illustrates the use of a named result object (Result). In the absence of a **return** statement specifying a result value, the final value of the named result object becomes the result of the operation.

### 2.3.4 Hash_Table, Keyed, and Key_Value Modules

Here are the interfaces for the Key_Value and Hash_Table modules used above in the Map module. Note that Hash_Table takes a single type parameter which must implement the Keyed interface, also given below. A Hash_Table is a good example of an irregular data structure, in that in a typical implementation, the number of elements in any given bucket of the hash table will vary depending on the number of keys that hash to the same hash-table index. Furthermore, when such a hash table hits some defined limit on density, the hash table is typically expanded to have more buckets, with the number of elements in each bucket again changing based on the enlarged range of the hashed index.

```
1  interface PSL::Containers::Hash_Table<KV_Type is Keyed<>> is
2      // A simple hash table of keyed entries
3      op "[]"() -> Hash_Table // Empty table
4      op "|="(var Left: Hash_Table; Right: KV_Type)
5          // Add Keyed entry to table
6      op "in" (Key: KV_Type::Key_Type; T: Hash_Table) -> Boolean
7          // True if Key in table
8      op "-="(var T: Hash_Table; Key: KV_Type::Key_Type)
9          // Remove entry with given key, if any
10     op "index_set"(T: Hash_Table) -> Set<KV_Type::Key_Type>
11         // Return set of keys in table
12     op "indexing"(ref T: Hash_Table; Key: KV_Type::Key_Type){Key in T} -> ref KV_Type
13         // Used for references to T[Key];
14     func Remove_Any(var T: Hash_Table) -> optional KV_Type
15         // Remove entry from table;
16         // Return null if table is empty.
17     op Count(T: Hash_Table) -> Univ_Integer
18         // Number of entries in the table
19 end interface PSL::Containers::Hash_Table
```

The interface Keyed requires only a single operation, Key_Of, as follows:

```
1  abstract interface Keyed<Key_Type is Hashable<>> is
2      func Key_Of(Keyed) -> Key_Type
3  end abstract interface Keyed
```

The Key_Value module defines two visible components (Key and Value), and provides the Key_Of operation, among others. It explicitly implements the Keyed interface.

ParaSail: A Pointer-Free Pervasively-Parallel LanguageBecause the Keyed interface has a parameter, the Key_Value interface must list it explicitly in the ancestry part of the interface declaration, to indicate how the formal types of Key_Value relate to the formal type of the Keyed interface:

```
1  interface Key_Value<Key_Type is Hashable<>; Value_Type is Assignable<>>
2    implements Keyed<Key_Type> is
3      var Key: Key_Type
4      var Value: Value_Type
5      func Key_Of(Key_Value) -> Key_Type
6      . . .
7  end interface Key_Value
```

Note that operation parameters may be specified merely by their type, in which case the name of the parameter defaults to the simple name of the type. In the above, the Key_Value parameter of Key_Of has both name Key_Value and type Key_Value. The distinction is resolved by context of usage. This short-hand is only permitted if each such default-named parameter has a distinct type.

Finally, here is the class part of the Hash_Table module, showing the underlying representation of the Hash_Table as an array of optional Nodes, each having a Keyed component and an optional Next component. Each node is effectively the head of a linked list, terminated by a node that has a null Next component. Note that in a language with no re-assignable pointers, it is useful to think of this not as a linked list, but rather as an expandable object, which grows somewhat like a vine, sprouting subcomponents out of the Next field (see section 3 below for more discussion of the pointer-free memory model). In this example, we see two internal data components, Backbone and Count, and a local module Node that has two visible components (Entry and Next) and no operations (thus requiring no class part to implement the module's interface). The functioning of the "in" and "index_set" operations is described in the next section.

```
1  class PSL::Containers::Hash_Table is
2      interface Node<> is
3          var Entry: Keyed
4          var Next: optional Node
5      end interface Node
6
7      var Backbone: Basic_Array<optional Node>
8      var Count: Univ_Integer:= 0
9    exports
10     op "[]"() -> Hash_Table
11         return (Backbone => [], Count => 0)
12     end op "[]"
13     . . .
14     op "in" (Key: KV_Type::Key_Type; T: Hash_Table) -> Boolean is
15         if T.Count > 0 then // non-empty
16             const H:= Hash(Key) mod |T.Backbone|
17             for N => T.Backbone[H+1]
18               then N.Next while N not null loop
19                 if Key_Of(N.Entry) == Key then
20                     return #true
21                 end if
```

```
22            end loop
23          end if
24          return #false
25       end op "in"
26       ...
27       op "index_set"(T: Hash_Table) -> Result: Set<KV_Type::Key_Type> is
28          // Build up set of keys
29          Result := []
30          for each B of T.Backbone loop
31             for N => B then N.Next
32                while N not null loop
33                   Result |= Key_Of(N)
34             end loop
35          end loop
36       end op "index_set"
37       op Count(T: Hash_Table) -> Univ_Integer
38          is (T.Count) // an expression function
39 end class PSL::Containers::Hash_Table
```

**Iterators and References** The Hash_Table class above illustrates a couple of operations with more complex implementations. The operator "in" hashes the given Key to identify the Hash_Table bucket which should be scanned to see whether the Key is already present. This uses one of the three forms of **for** loop provided by ParaSail, the one allowing an initial object (**for** N => Root) or value (**for** I := First), the next object or value (**then** N.Next or **then** I+1), and a termination test (**while** N **not null**, or **while** I <= Last).

The "index_set" operator returns a set of keys, given a hash table. This uses both the **for/then/while** form of **for** loop, as well as a *container element iterator*, which uses **for each** Element **of** Indexed_Container to iterate through the set of Elements of the given indexed container. The element iterator is syntactic sugar for the following expansion:

```
1    var @Keys := "index_set"(Indexed_Container)
2    for @K := Remove_Any(@Keys)
3      then Remove_Any(@Keys)
4      while @K not null loop
5        ref Element => Indexed_Container[@K]
6        ... // body of loop
7    end loop
```

where @K and @Keys are meant to represent compiler-generated unique names.

In the above expansion, and in the declaration for the "indexing" operators, we see the use of the **ref** keyword to indicate a local, short-lived name for an existing object (Element is the local name for the given element of Indexed_Container). Such names inherit the type and writability of the referenced object. Any use of such a ref is equivalent to a use of the referenced object. An operation may return a ref, but only if it is a reference to some part of an object passed to it as a ref parameter. This is how the "indexing" operator works.

**ParaSail: A Pointer-Free Pervasively-Parallel Language**

**2.4 ParaSail Literals and Univ Types**

In the above examples, we have used integer literals, such as 0 and 1, without describing their semantics. In addition we have used the Univ_Integer type and also Integer as an example of a type over which a Set might be defined. There are actually five different sorts of literals in ParaSail: integer, real, character, string, and enumeration. Each has a distinct syntax, and each has an associated type in which the literal's value is initially represented. This is summarized in the following table:

| *Kind of Literal* | *Example Syntax* | *Type* |
| --- | --- | --- |
| Integer | 123, 0xFF, 8#77# | Univ_Integer |
| Real | 123.4, 1.2E7 | Univ_Real |
| Character | 'i', '\n', '\#03_C0#' | Univ_Character |
| String | "abc", "two\nlines" | Univ_String |
| Enumeration | #red, #false, #equal | Univ_Enumeration |

Most of the ParaSail syntax for literals is quite similar to other languages such as C or Java. The '\#...#' form of character literal is used to specify the Unicode value of the literal. Enumeration literals have a distinct syntax in ParaSail. This is common in Lisp-based languages, but less common in imperative languages. By using a distinct syntax, with its own Univ_Enumeration type, we enable significant flexibility to users in defining the mapping from such literals to the values of a given type.

In general, to allow a given kind of literal to be used with a user-defined type, the associated module must define two operators, "from_univ" and "to_univ". These are conversion functions, from a Univ type, or back to such a type. By defining a function of the form `"from_univ"(Univ_Integer) -> My_Type` integer literals may be used to represent values of My_Type. Similarly, by defining a `"from_univ"(Univ_Enumeration) -> My_Enum_Type`, enumeration literals may be used to represent values of My_Enum_Type. Calls on these conversion functions are inserted implicitly during name and type resolution by the ParaSail front end, when a literal is used within an expression. Effectively, a literal is actually syntactic sugar for "from_univ"(literal).

These "from_univ" conversion functions must include a precondition (specified with the syntax "{Boolean-expression}" as illustrated in the "indexing" operators above) which determines which literals can be converted into values of the given type. So, for example, the precondition for the Ordering type's "from_univ" operator, presuming it is declared `"from_univ"(Univ : Univ_Enumeration) -> Ordering`, would be:

   *{Univ in [#less,#equal,#greater,#unordered]}*

By specifying such a precondition, the compiler will disallow use of other literals of the Univ_Enumeration type when it expects a value of type Ordering.

**2.4.1 User-Defined Numeric and Enumeration Types**

Values and objects of the Univ types can in fact be used at run-time as well, but their representation and manipulation will generally not be as efficient as a user-defined type that need only represent some particular range of values. In particular, in the ParaSail standard library there are modules designed for more efficient representation of numeric and enumeration types, including an Integer, Float, Fixed, and Enum module. The Integer module defines all of the usual operators, and has one module parameter which is the range of Univ_Integers that are to be representable in objects of the type:

```
1  interface Integer <Range: Countable_Range<Univ_Integer> := -2**63+1 .. +2**63-1> is
2      op "from_univ" (Univ : Univ_Integer) {Univ in Range} -> Integer
3      op "to_univ" (Integer) -> Result : Univ_Integer {Result in Range} // postcondition
4      op "+" (Left, Right : Integer) -> Integer
5      ...
6  end interface Integer
```

Because the Range parameter has a default, Integer by itself is interpreted as an instantiation of the module Integer with all parameters defaulted, i.e. Integer is a short-hand for Integer<> which is a short-hand for Integer<Range => -2**63+1 .. +2**63-1>. Note that ".." is a user-definable operator in ParaSail, and is expected to return a set or a range of values. Also note that "from_univ" has a *precondition*, while "to_univ" has a *postcondition*, based on the specified Range.

The Univ types are universal in the sense that values of a Univ type are implicitly convertible to and from all types of the corresponding kind. So in addition to the implicit calls on "from_univ" applied to literals, if a formal parameter of some operation is declared to be of a Univ type, then a call on the appropriate "to_univ" conversion will be inserted as appropriate if an operand of a non-universal type of the corresponding kind is passed as the actual parameter. For example, a formal parameter of type Univ_Integer accepts *any* integer type as its actual parameter. So Univ types effectively allow *weak* typing, while non-Univ types enforce *strong* typing for numeric, character, string, and enumeration types.

## 3 Pointer-Free Region-Based Storage

Pointers are ubiquitous in modern object-oriented programming languages, and many data structures such as trees, lists, graphs, hash tables, etc. depend on them heavily. Unfortunately, pointers can add significant complexity to programming. Pointers tend to make storage management more complex, to make assignment and equality semantics more complex, to increase the ways two different names (access paths) can designate the same object, to make program analysis and proof more complex, and to make it harder to partition a data structure for *divide-and-conquer* parallel processing.

### 3.1 Expandable and Optional Objects

Rather than using pointers, ParaSail supports flexible data structuring using *expandable* (and shrinkable) objects, along with generalized indexing. An expandable object is one that can grow without using pointers, much as a plant can grow through sprouting new stems. The basic mechanism for expansion in ParaSail is as mentioned in section 2.2 above, namely that every type has one additional value, called null. A component can initially be null, and then be replaced by a non-null value, thereby expanding the enclosing object. At some later point the enclosing object could shrink, by replacing a non-null component with null. Expandable objects are managed efficiently using a variant of region-based storage management (see section 3.3 below).

As also mentioned in section 2.2, not every component of an object is allowed to be null. The component must be declared as **optional** if it is allowed to take on a null value. For example, a Tree structure might have a (non-optional) Payload component, and then two additional components, Left and Right, which are each declared as **optional** Tree. Similarly, a stand-alone object may be declared to be of a type T, or of a type **optional** T. Only if it is declared optional may it take on the null value. The value of an object X declared as optional may be tested for nullness using X **is null** or X **not null**.

Another example of a data structure using optional components would be a linked list, with each node having two components, say, a Payload component of some type, and a Tail component of type **optional** List. There is also a built-in parameterized type, Basic_Array<Component_Type> which allows the Component_Type to be specified as optional. This allows, for example, the construction of a hash table with buckets represented as linked-lists, by declaring the backbone of the hash table as a Basic_Array<**optional** Node> as illustrated in the Hash_Table example in section 2.3.4 above. The elements of the hash table backbone would start out as null, but as items are added to the hash table, one or more of the component lists would begin to grow. In this case, a Node is defined as having an Entry component and an optional Next component of type Node.

### 3.1.1 Assignment, Move, and Swap Operations

Because there are no pointers, the semantics of *assignment* in ParaSail are straightforward, namely the entire right-hand-side object is copied and assigned into the left-hand side, replacing whatever prior value was there. However, there are times when it is desirable to move a component from one object to another, or swap two components. Because implementing these on top of an assignment that uses copying might impose undue overhead, in ParaSail, *move* and *swap* are separate operations. The semantics of *move* are that the value of the left-hand-side is replaced with the value of the right-hand-side, and the right-hand-side ends up null. For *swap*, the values of the left- and right-hand-side are swapped. Syntactically, ParaSail uses ":=" for (copying) assignment, "<==" for move, and "<=>" for swap. The ParaSail compiler is smart enough to automatically use move semantics when the right-hand-side is the result of a computation, rather than an object or component that persists after the assignment.

As an example of where move might be used, if our hash table grows to the point that it would be wise to lengthen the backbone, we could create a new Basic_Array twice as large (for example), and then move each list node from the old array into the new array in an appropriate spot, rebuilding each linked list, and then finally move the new array into the original hash-table object, replacing the old array. Here is sample code for such an expansion:

```
1    func Expand (var HT : Hash_Table) is
2       // Double the size of the Hash_Table backbone
```

```
3        var Old_Backbone <== HT.Backbone // Move old backbone to temp
4        const New_Len := |Old_Backbone| * 2
5        // Create backbone with double the number of buckets
6        HT.Backbone := Create(New_Len, null)
7        for each Old_Bucket of Old_Backbone loop
8           for Old_Elem => Old_Bucket then Old_Elem.Next while Old_Elem not null loop
9              const New_Hash := Hash(Key_Of(Old_Elem.Entry)) mod New_Len
10             ref New_Bucket => HT.Backbone[New_Hash + 1]
11             // Insert at front, using ``move''s to create new node.
12             New_Bucket := Node::(Entry <== Old_Elem.Entry, Next <== New_Bucket)
13          end loop
14       end loop
15    end func Expand
```

The swap operation is also useful in many contexts, for example when balancing a tree structure, or when sorting an array.

### 3.2 Cyclic Data Structures and Generalized Indexing

Expandable objects allow the construction of many kinds of data structures, but a general, possibly cyclic graph is not one of them. For this, ParaSail provides generalized indexing. The array-indexing syntax, A[I], is generalized in ParaSail to be usable with any container-like data structure, where A is the container and I is the key into that data structure. A directed graph in ParaSail could be represented as a table of Nodes, where the index into the table is a unique Node Id of some sort, with edges represented as Predecessors and Successors components of each Node, where Predecessors and Successors are each sets of node-ids.

If edges in a directed graph were instead represented with pointers, it would be possible for there to be an edge that refers to a deleted node, that is, a dangling reference. Such a dangling reference could result in a storage leak, because the target node could not be reclaimed, or it could lead to a potentially destructive reference to reclaimed storage. By contrast, when edges are represented using node-ids, there is still the possibility of an edge referring to a deleted node or the wrong node, but there is no possibility for there to be associated storage leakage or destructive reference to reclaimed storage, as node-ids are only meaningful as keys into the associated container.

### 3.3 Region-Based Storage Management

Storage management without pointers is significantly simplified, even with highly irregular data structures. All of the objects declared in a given scope are associated with a *storage region*, essentially a local heap. As an object grows, all new storage for it is allocated out of this region. As an object shrinks, the old storage can be immediately released back to this region. When a scope is exited, the entire region is reclaimed. There is no need for asynchronous garbage collection, as garbage never accumulates. Objects may grow in a highly irregular fashion without losing their locality of reference.

Every object identifies its region, and in addition, when a function is called, the region in which the result object should be allocated is passed as an implicit parameter. This target region is determined by how the function result is used. If it is a temporary, then it will be allocated out of a temporary region associated with the point of call. If it is assigned into a longer-lived object, then the function will be directed to allocate the result object out of the region associated with this longer-lived object. The net effect is that there is no copying at the call site upon function return, since the result object is already sitting in the correct region.

Note that pointers are still used behind the scenes in the ParaSail implementation, but eliminating them from the surface syntax and semantics eliminates the complexity associated with pointers. That is, a semantic model of expandable and shrinkable objects, operating under (mutable) value semantics, rather than a semantic model of nodes connected with pointers, operating under reference semantics, provides a number of benefits, such as simpler storage management, simpler assignment semantics, easier analyzability, etc. while preserving flexibility in representing potentially highly irregular structures.

ParaSail's move and swap operations have well-defined semantics independent of the region-based storage management, but they provide significant added efficiency when the objects named on the left and right-hand side are associated with the same region, because then their dynamic semantics can be accomplished simply by manipulating pointers. In some cases the programmer knows when declaring an object that it is intended to be moved into or swapped with another existing object. In that case, ParaSail allows the programmer to give a hint to that effect by specifying in the object's declaration that it is "**for** X" meaning that it should be associated with the same region as X. With region-based storage management, it is always safe to associate an object with a longer-lived region, but to avoid a storage leak the ParaSail implementation sets the value of such an object to null on scope exit, as its storage would not otherwise be reclaimed until the longer-lived region is reclaimed. An optimizing compiler could automatically choose to allocate a local variable out of an outer region when it determines that its last use is a move or an assignment to an object from an outer region.

It is straightforward to show that when a non-null object or a component is set to null, immediate reclamation of the old value is possible, and will not create a dangling reference. Assignment copies rather than shares data, and move and swap create no sharing relationships, so each piece of storage used to represent the value of an object is unshared.

One possible way to create a dangling reference might be if two computations were being performed in parallel, and one were to set an object to null while the other was still using the object. The mechanism to prevent this is part of the more general mechanism to prevent concurrent update of an object while other computations have access to it. This is covered in the next section.

## 4 Parallel and Distributed Programming

In addition to removing pointers, certain other simplifications are made in ParaSail to ease parallel and distributed programming of potentially irregular computations. In particular, there are no global variables; functions may only update objects passed to them as **var** (in-out) parameters. Furthermore, as part of passing an object as a **var** parameter, it is effectively *handed off* to the receiving function, and compile-time checks ensure that no further references are made to the object, until the function completes. In particular, the checks ensure that no part of the **var** parameter is passed to any other function, nor to this same function as a separate parameter. This eliminates at compile time the possibility of aliasing between a **var** parameter and any other object visible to the function. These two additional rules, coupled with the lack of pointers, mean that all parameter evaluation may happen in parallel (e.g. in G(X) + H(Y), the operands to "+", G(X) and H(Y), may be evaluated in parallel). These rules also imply that a ParaSail function call can safely cross an address-space boundary, since the objects are self-contained (with no incoming or outgoing references), and only one function at a time can update a given object.

Because of the simplifications in ParaSail, conservative data race detection can be incorporated directly into the compiler front end, ensuring that no data races remain in the program independent of whether expressions are evaluated sequentially or in parallel. This data race detection permits the compiler to safely and automatically insert parallel evaluation even across sequential statements in a ParaSail programs, so long as the compiler can determine it would not introduce a data race. Furthermore, in ParaSail the programmer can explicitly claim that two statements can safely be executed in parallel, by using "||" rather than ";" as the separator between the statements. Alternatively, the programmer can specify that two statements must not be executed in parallel, by using **then** rather than ";" as the separator.

As a simple example of an irregular computation that uses parallelism, below is an exhaustive search of a binary tree, where each Tree_Node has four components, a Key (of type Key_Type) a Value (of type Value_Type), and Left and Right subtrees (each of type **optional** Tree_Node). Each iteration of the loop checks to see whether the Value of the node referred to by T matches Desired_Value. If so, it returns the associated Key field of the node, and the function ends. Otherwise, it uses parallel **continue loop** statements to spawn two more iterations, one to search the Left subtree, and one to search the Right subtree. The **while** T **not null** of the loop header acts as a filter on the iterations, immediately terminating an iteration if T is null. If T is not null, the iteration proceeds, and either returns or spawns two more iterations. A loop like this continues until all iterations complete (in this example, the function would then return null), or until one of the iterations does a return, in which case all other iterations are terminated, and then the return causes the function to exit with the given value. If two or more iterations concurrently reach a return (or exit) statement, one is chosen arbitrarily to proceed while the others are terminated. This is effectively viewing a loop as a bag of iterations executing in parallel, with **continue loop** adding another iteration to the bag, and the loop terminating when the bag of iterations is empty, or when one (or more) of the iterations does a return (or exit).

ParaSail: A Pointer-Free Pervasively-Parallel Language```
1    func Search(Root : Tree_Node; Desired_Value : Value_Type) -> optional Key_Type is
2       for T => Root while T not null loop
3          if T.Value == Desired_Value then
4             return T.Key
5          else
6             continue loop with T.Left
7             ||
8             continue loop with T.Right
9          end if
10      end loop
11      return null
12   end func Search
```

When one iteration forces another iteration to terminate, the iteration to be terminated is flagged, and the underlying scheduler attempts to terminate the iteration as soon as it is safe and efficient to do so, where "safe" in this context means that the flagged iteration is not currently executing within the body of a **locked** (or **queued**) operation of a **concurrent** object (see section 4.2 below), and "efficient" means that the implementation may choose to provide immediate termination, or instead to check for such a termination flag only periodically. In any case, any attempt by a flagged iteration to spawn a new parallel computation (outside of a locked operation) results in immediate termination. An additional ParaSail safety rule disallows updating a non-**concurrent** variable declared outside a concurrent loop from within the loop, if the loop has a statement that might result in early termination, such as a return or exit. This ensures that early termination of an iteration will not disrupt an update of a (non-concurrent) variable that outlives the loop.

A more complete example of the parallelism features is provided in the next section, by a parallel, non-recursive implementation of in-place Quicksort.

### 4.1 Parallel Non-Recursive Quicksort

Quicksort is an example of an irregular computation, in that it repeatedly partitions the array to be sorted at points that depend both on the values in the array and on which values are chosen as pivots. Quicksort also illustrates the use of swap as the primary mutation operation on the array being sorted in place. Below is an in-place version of Quicksort that uses parallelism in a number of ways, with swap as the fundamental data-moving operation. The slicing operation A[X .. Y] produces a slice of the array A comprising the elements with indices from X through Y. The operation A[..] is used to create a full slice that goes from the first to the last element of A, essentially viewing the array as a slice.

The Quicksort operation is structured as an outer loop where each iteration sorts a slice of the array. The first iteration operates on a full slice of the array. Additional iterations are created by **continue loop** statements on sub-slices of the array. As illustrated above in the Search example, when two continue statements are invoked in parallel, then the two iterations can run in parallel with one another. The loop as a whole completes once all of the iterations of the loop are complete. Effectively this sort of





ParaSail loop implements a kind of work-list algorithm, with **continue loop** being the way to add a new work-item to the work-list. Note that similar to the Search example, the **while** |Arr| > 1 acts as a filter on the iterations, in that iterations for a slice of length <= 1 terminate immediately.

In each iteration, if the array-slice Arr is of length exactly two, then the two elements are checked to see whether they are in order, and if not, they are swapped (using the ParaSail "<=>" swap operation). If the slice Arr comprises more than two elements, then the iteration proceeds to partition the slice into subslices, by picking a pivot value (Mid), and then looking (in parallel) for two elements that are on the "wrong" side, and swapping them (again using "<=>"). Once there are no more elements to swap, the original array slice has been partitioned, and the two partitions (subslices) are themselves sorted. Rather than recursion, the implicit bag or work-list model of this kind of ParaSail **for** loop allows a non-recursive solution, with each subslice to be sorted merely being added to the bag or work-list of iterations to be performed, rather than resulting in a recursion.

```
1    func Quicksort(var A : Array_Type) is
2        // Handle short arrays directly. Partition longer arrays.
3        for Arr => A[..] while |Arr| > 1 loop
4            if |Arr| == 2 then
5                if Arr[Arr.Last] < Arr[Arr.First] then
6                    // Swap elements
7                    Arr[Arr.First] <=> Arr[Arr.Last];
8                end if;
9            else
10               // Partition array
11               const Mid := Arr[Arr.First + |Arr|/2];
12               var Left := Arr.First;
13               var Right := Arr.Last;
14               until Left > Right loop
15                   var New_Left := Right+1;
16                   var New_Right := Left-1;
17                 then
18                   // Find item in left half to swap
19                   for I in Left .. Right forward loop
20                       if not (Arr[I] < Mid) then
21                           // Found an item that can go into right partitition
22                           New_Left := I;
23                           if Mid < Arr[I] then
24                               // Found an item that *must* go into right part
25                               exit loop;
26                           end if;
27                       end if;
28                   end loop;
29                 ||
30                   // Find item in right half to swap
31                   for J in Left .. Right reverse loop
32                       if not (Mid < Arr[J]) then
33                           // Found an item that can go into left partition
34                           New_Right := J;
```





```
35                              if Arr[J] < Mid then
36                                  // Found an item that *must* go into left part
37                                  exit loop;
38                              end if;
39                          end if;
40                      end loop;
41                  then
42                      if New_Left > New_Right then
43                          // Nothing more to swap
44                          // Exit loop and recurse on two partitions
45                          Left := New_Left;
46                          Right := New_Right;
47                          exit loop;
48                      end if;
49                      // Swap items
50                      Arr[New_Left] <=> Arr[New_Right];
51                      // continue looking for items to swap
52                      Left := New_Left + 1;
53                      Right := New_Right - 1;
54                  end loop;
55                then
56                  // continue with two halves in parallel
57                  continue loop with Arr => Arr[Arr.First .. Right];
58                ||
59                  continue loop with Arr => Arr[Left .. Arr.Last];
60              end if;
61          end loop;
62      end func Quicksort;
```

It is worth noting that ParaSail also has sufficient expressiveness that a much simpler, fully "functional" implementation of Quicksort is possible, though it is no longer solving the challenge involved in an in-place sort provided by the above. Here is such a fully functional (non in-place) implementation of recursive Quicksort in ParaSail, using an extensible array-like generic Vector abstraction over Comparable components:

```
1   func Qsort(V : Vec_Type is Vector<Comparable<>>) -> Vec_Type is
2       if |V| <= 1 then
3           return V; // The easy case
4       else
5           const Mid := V[ |V|/2 ]; // Pick a pivot value
6           return
7               QSort( [for each E of V {E < Mid} => E] ) // Recurse
8               | [for each E of V {E == Mid} => E] // No recursion since all values equal the pivot
9               | QSort( [for each E of V {E > Mid} => E] ); // Recurse
10      end if;
11  end func Qsort;
```

The above constructs three sub-vectors from the elements of the original vector V using filters E < Mid, E == Mid, and E > Mid specified in braces, and then concatenates the sorted sub-vectors using the | operator. ParaSail's default concurrent evaluation of complex operands of a binary operation will produce similar levels of parallelism.





## 4.2 Concurrent Objects

The *handoff* model of parameter passing applies to objects that are not designed for concurrent access. ParaSail also supports the construction of *concurrent* objects, which allow lock-free, locked, and queued simultaneous access. Concurrent objects are not "handed off" as part of parameter passing, and aliasing of such parameters is permitted; concurrent objects provide operations that synchronize any attempts at concurrent access. Three kinds of synchronization are supported. Lock-free synchronization relies on low-level hardware-supported operations such as atomic load and store, and compare-and-swap. Locked synchronization relies on automatic locking as part of calling a locked operation of a concurrent object, and automatic unlocking as part of returning from the operation. Finally, queued synchronization is provided, which evaluates a *dequeue* condition upon call (under a lock), and only if the condition is satisfied is the call allowed to proceed, still under the lock. A typical dequeue condition might be that a buffer is not full, or that a mailbox has at least one element in it. If the dequeue condition is not satisfied, then the caller is added to a queue. At the end of any operation on the concurrent object that might change the result of the dequeue condition for a queued caller, the dequeue condition is evaluated and if satisfied, the operation requested by the queued caller is performed before the lock is released. If there are multiple queued callers, then they are serviced in turn until there are none with satisfied dequeue conditions.

One way to understand the distinction between "normal" objects and concurrent objects is that the compiler performs *compile-time* checks to ensure there are no data races on accessing "normal" objects, while concurrent objects use *run-time* synchronization to prevent data races.

Below is a simple example of a concurrent module, a `Locked_Box`. This concurrent module has both an interface and a class that implements it. In the interface we see the declaration of five operations: one operation that constructs a `Locked_Box` (`Create`), one locking operation that overwrites the content (`Set_Content` which gets an exclusive read-write lock on its **var** parameter B), one locking operation that reads the content (`Content` which gets a shared lock on its read-only parameter B), and two queuing operations, one that will add a value to an empty Box (`Put`), and one that will remove a value from a full Box (`Get`). In the class for the module `Locked_Box`, we see the internal data component Content which is declared optional to indicate that it can be null, followed by the implementations of the five operations. The operations that have locked or queued access can be written knowing that an appropriate exclusive or shared lock is acquired as part of the call of the operation. Furthermore, for the queued operations, the call is queued until the specified dequeue condition is satisfied. In this case, the dequeue condition for Put ensures the box is empty (i.e. has a null value) and the dequeue condition for Get ensures the box is full (i.e. has a non-null value). Note that a queued operation can safely assume that the dequeue condition is satisfied when it begins. ParaSail semantics ensure there is no need to recheck the condition explicitly. The locking, waiting, and signaling to support the semantics of these operations are all provided automatically in ParaSail as part of the semantics of a call on such an operation.



ParaSail: A Pointer-Free Pervasively-Parallel Language

```
 1  concurrent interface Locked_Box <Content_Type is Assignable<>> is
 2      func Create(C : optional Content_Type) -> Locked_Box;
 3          // Create a box with the given content
 4      func Set_Content (locked var B : Locked_Box; C : optional Content_Type);
 5          // Set content of box
 6      func Content(locked B : Locked_Box) -> optional Content_Type;
 7          // Get a copy of current content
 8      func Put(queued var B : Locked_Box; C : Content_Type);
 9          // Wait for the box to be empty, and then Put something into it.
10      func Get(queued var B : Locked_Box) -> Content_Type;
11          // Wait until content is non-null, then return it, leaving it null.
12  end interface Locked_Box;
13
14  concurrent class Locked_Box is
15      var Content : optional Content_Type; // Content might be null
16    exports
17      func Create(C : optional Content_Type) -> Locked_Box is
18          // Create a box with the given content
19          return (Content => C);
20      end func Create;
21
22      func Set_Content (locked var B : Locked_Box; C : optional Content_Type) is
23          // Set content of box
24          B.Content := C;
25      end func Set_Content;
26
27      func Content(locked B : Locked_Box) -> optional Content_Type is
28          // Get a copy of current content
29          return B.Content;
30      end func Content;
31
32      func Put(queued var B : Locked_Box; C : Content_Type) is
33        queued until B.Content is null then
34          // Wait for the box to be empty,
35          // and then Put something into it.
36          B.Content := C;
37      end func Put;
38
39      func Get(queued var B : Locked_Box) -> Result : Content_Type is
40        queued while B.Content is null then
41          // Wait until content is non-null,
42          // then return it, leaving it null.
43          Result <== B.Content;
44      end func Get;xs
45  end class Locked_Box;
```

The above Get operation makes use of the move construct ("<==") which moves the value of the right-hand side (B.Content) into the left-hand side (Result) leaving the right-hand side null afterward. This matches the desired semantics, namely that Get waits until the box B is "full" (i.e. non-null), but then leaves it "empty" (i.e. null) upon return.





## 5 Related Work

There are very few pointer-free languages currently under active development. Fortran 77 [41] was the last of the Fortran series that restricted itself to a pointer-free model of programming. Algol 60 lacked pointers [2], but Algol 68 introduced them [22]. Early versions of Basic had no pointers [19], but modern versions of Basic use pointer assignment semantics for most complex objects [25]. The first versions of Pascal, Ada, Modula, C, and C++ all used pointers for objects that were explicitly allocated on the heap, while still supporting stack-based records and arrays; these languages also required manual heap storage reclamation. The first versions of Eiffel, Java, and C# provided little or no support for stack-based records and arrays, moving essentially all complex objects into the heap, with pointer semantics on assignment, and automatic garbage collection used for heap storage reclamation.

In many cases, in languages that originally did not require heavy use of pointers, as they evolved to support object-oriented programming, the use of pointers increased, often accompanied by a reliance on garbage collection for heap storage reclamation. For example, Modula-3 introduced object types, and all instances of such types were allocated explicitly on the heap, with pointer semantics on assignment, and automatic garbage collection for storage reclamation [8].

The Hermes language (and its predecessor NIL) was a language specifically designed for distributed processing [32]. The Hermes type system had high-level type constructors, which allowed them to eliminate pointers. As the designer of Hermes explained it, "pointers are useful constructs for implementing many different data structures, but they also introduce aliasing and increase the complexity of program analysis" [32, p. 80]. NIL, the precursor to Hermes, pioneered the notion of *type state* [33], as well as *handoff* semantics for communication, both of which are relevant to ParaSail, where compile-time assertion checking depends on flow analysis, and handoff semantics are used for passing **var** parameters in a call on an operation.

Another distributed-systems language that is pointer-free is Composita, described in the 2007 Ph. D. thesis of Dr. Luc Bläser from ETH in Zurich [4]. Composita is a component-based language, which uses message passing between active components. Sequences of statements are identified as either exclusive or shared to provide synchronization between concurrent activities. Composita has the notion of empty and installed components, analogous to the notion of optional values in ParaSail.

The SPARK language, a high-integrity subset of Ada with added proof annotations, omits pointers from the subset [9]. No particular attempt was made to soften the effect of losing pointers, so designing semi-dynamic data structures such as trees and linked-lists in SPARK requires heavy use of arrays [36, 37].

Annotations that indicate an ownership relationship between a pointer and an object can provide some of the same benefits as eliminating pointers [10]. AliasJava [1] provides annotations for specifying ownership relationships, including the notion of a unique pointer to an object. Guava [3] is another Java-based language that adds value types which have no aliases, while still retaining normal object types for other purposes. Assignment of value types in Guava involves copying, but they also provide a move operation essentially equivalent to that in ParaSail. Related work by Boyapati





and Rinard [6, 7] on Parameterized Race-Free Java focuses on the elimination of data races through the use of thread-based ownership, augmented with transferable unique pointers. These approaches, by using ownership and unique pointers to limit the possibilities for aliasing, can significantly help in proving desirable properties about programs that use pointers. However, the additional programmer burden of choosing between multiple kinds of pointers or objects based on their ownership or aliasing behavior can increase the complexity of such approaches.

One reason given in these papers on aliasing control for not going entirely to a pointer-free, or unique-pointer approach for object-oriented programming, is that certain important object-oriented programming paradigms, such as the Observer pattern [12], depend on the use of pointers and aliasing. ParaSail attempts to provide an existence proof to the contrary of that premise, as do other pointer-free languages such as Composita and Hermes. In general, a more loosely-coupled pointer-free approach using container data structures with indices of various sorts, allows the same problem to be solved, with fewer storage management and synchronization issues. For example, the Observer pattern, which is typically based on lists of pointers to observing objects, might be implemented using a pointer-free Publish-Subscribe pattern, which can provide better scalability and easier use of concurrency [17]. In general, pointers are not directly usable in distributed systems, so many of the algorithms adopted to solve problems in a distributed manner are naturally pointer-free, and hence are directly implementable in ParaSail.

The Rust language [28] from Mozilla Research adopted unique pointers for all inter-task communication, thereby eliminating all race conditions between tasks. Originally the designers of Rust intended to allow non-unique pointers for intra-task operations, but in subsequent writings [40] they argued that unique pointers can provide adequate flexibility, thus providing more evidence that fully general pointers, with the attendant need for garbage collection, are not needed even for complex data structure manipulations. More generally, the Rust language addresses many of the same challenges that ParaSail addresses using similar approaches, such as a *borrowing* mechanism on parameter passing, analogous to the *handoff* semantics used in ParaSail. Rust goes somewhat beyond the implicit semantics of ParaSail's handoff semantics and region-based storage management, providing, for example, the ability to annotate references with explicit lifetimes [29].

Object ownership continues to be an area of active research, driven by interest in safe parallel programming, as well as the efficiency and predictability of garbage-collection-free storage management. The LaCasa system developed by Haller and Loiko [14], builds on standard Scala extension mechanisms, with additional checking provided by a compiler plug-in, to enforce an *object-capability discipline* that essentially disallows direct reference to global variable state, requiring all references to such state to be mediated by an *access permission* and an explicit *open* operation. This approach enforces a Scala variant that closely matches the fundamental model provided by ParaSail, where there are no global variables, and where functions may only access objects passed to them as parameters. Project Snowflake [30] adds to the Microsoft .NET framework support for manual memory management of a portion of the heap storage, using an approach that ensures memory and thread safety through





a combination of ownership, and thread-specific *shields*. A shield prevents an object in use by a thread from being reclaimed until the shield is removed. In both LaCasa and Project Snowflake, because they are built upon an existing language, they cannot adopt uniform *handoff* or *borrowing* semantics on parameter passing, and instead depend on an extra step such as opening a box mediated by a permission, or initializing a shield mediated by a unique owning reference, to gain safe, exclusive access to the content of the owned object. Also, both forsake the added flexibility of *concurrent read, exclusive write* [31] provided by ParaSail and Rust, and limit themselves to exclusive access, whether read or write, to limit the burden of additional user annotations while still ensuring thread safety.

Pure functional languages, such as Haskell [23], avoid many of the issues of pointers by adopting immutable objects, meaning that sharing of data creates no aliasing or race condition problems. However, mostly functional languages, such as those derived from the ML language [15], include references to mutable objects, thereby re-introducing most of the potential issues with aliasing and race conditions. Even Haskell has found it necessary to introduce special monads such as the IO monad to support applications where side-effects are essential to the operation of the program. In such cases, these side-effects need to be managed in the context of parallel programming [24].

Hoare in his 1975 paper on Recursive Data Structures [16] identified many of the problems with general pointers, and proposed a notation for defining and manipulating recursive data structures without the use of pointers at the language level, even though pointers were expected to be used at the implementation level. Language-level syntax and semantics reminiscent of this early proposal have appeared in functional languages, but have not been widely followed in languages with mutable values. Mostly-functional languages such as ML have also more followed the Algol 68 model of explicit references when defining mutable recursive data structures, despite Hoare's many good arguments favoring a pointer-free semantics at the language level. Hoare's notation did not introduce the notion of optional values, but instead relied on types defined by a tagged union of generators, at least one of which was required to be non-recursive. ParaSail adopts the optional value approach and allows the set of generators that can be used to create objects to be open-ended, by relying on object-oriented polymorphism over interfaces.

Minimizing use of a global heap through the use of region-based storage management was proposed by Tofte and Talpin [38] and implemented in the ML Kit with Regions [39]. Region-based storage management was brought to a safe subset of C in the language Cyclone [13]. Cyclone, being a derivative of C, was not a pointer-free language. Instead, every pointer was associated with a particular region at compile time, allowing compile-time detection of dangling references. A global, garbage-collected heap was available, but local dynamic regions provided a safe, more efficient alternative. In both the ML Kit with Regions and in Cyclone, region inference was performed to minimize the need for explicit region annotations.

More recent work by Kaki and Ramalingam [18] has linked the notion of region-based storage management with transferable ownership, with regions as owners, and with subobjects being generally owned by the same region as their containing object, while allowing an entire region to be safely transferred from one address space to





another as part of message passing. No particular attempt is made to ensure thread safety through this region-based ownership – the focus is strictly on memory safety.

Many functional (or mostly functional) languages have a notion similar to ParaSail's optional objects. For example, in Haskell they are called *maybe* objects [23]. In ParaSail, because of its fundamental role in supporting recursive data structures, **optional** is a built-in property usable with every object, component, or type declaration, rather than being an additional level of type. In addition, this approach allows null-ness to be represented without a distinct null object, by ensuring that every type has at least one bit pattern than can be recognizable as a null value.

## 6 Implementation Status and Evaluation

A version of the ParaSail front end and accompanying documentation is available for download [34]. The front end generates instructions for a ParaSail Virtual Machine (PSVM). A full multi-threaded interpreter for the PSVM instruction set is built into the front end, and includes a simple interactive Read-Eval-Print Loop for testing. A backend that translates from the PSVM instruction set to LLVM has been developed, along with a static analysis tool to allow the compiler to enforce preconditions, postconditions, and nullness checks at compile time.

The ParaSail front end automatically splits computations up into very light-weight picothreads, each representing a potentially parallel sub-computation. The PSVM includes special instructions for spawning and awaiting such picothreads. The PSVM run-time uses the *work stealing* model [5] to execute the picothreads; work stealing incorporates heavier weight server processes which each service their own queue of picothreads (in a LIFO manner), stealing from another server's queue (in a FIFO manner) only when their own queue becomes empty.

ParaSail adopted a pointer-free model initially to enable easy and safe pervasively parallel programming for potentially irregular computations. However, the ongoing experience in programming in ParaSail with its pointer-free, mutable value semantics, has provided support for the view that pointers are an unnecessary burden on object-oriented programming. The availability of optional values allows the direct representation of tree structures, singly-linked lists, hash tables, and so on in much the same way they are represented with pointers, but without the added complexities of analysis, storage management, and parallelization associated with pointers.

Data structures that require multiple paths to the same object, such as doubly-linked lists or general graphs, can be implemented without pointers by using indexing into generalized container structures. Even in languages without the restriction against pointers, it is not uncommon to represent directed graphs using indices rather than pointers, in part because the presence or absence of edges between nodes does not necessarily affect whether the node itself should exist. An additional advantage to using a container such as a vector to represent a graph, is that partitioning of the graph for the purpose of a parallel divide-and-conquer computation over the graph can be simplified, by using a simple numeric range test on the index to determine whether a given node is within the subgraph associated with a particular sub-computation.





Operations on indices tend to be easier to analyze than those on pointers, including, for example, a proof that two variables contain different indices, as would be needed for a proof of non-aliasing when the indices are used to index into a container.

Evaluating a programming language in a fully objective sense is challenging, and even deciding on what criteria are relevant is by itself difficult. The "-ilities" and "-arities" such as usability, reliability, portability, modularity, etc., might be easily agreed to be relevant, but ranking languages according to such criteria generally involves subjective evaluation, bordering on the "religious" at times. For the purposes of evaluating the design of ParaSail, we have attempted to assess how well we accomplished our various goals, and how ParaSail compares to other modern systems programming languages in achieving these goals.

One of our initial goals for ParaSail was to create a language that would be familiar to existing professional systems programmers, to ease comprehension and adoption. Even though at its core ParaSail adopts quite different fundamental semantic principles in terms of parallelism and data structuring, nevertheless we have attempted to make its user-level syntax and semantics something that looks and feels similar to existing strongly-typed, object-oriented languages. In our informal experiments, programmers familiar with Java or C# have not had trouble understanding and learning to program potentially irregular applications in ParaSail, thanks in part to its familiar class-and-interface object-oriented programming model. The notion of optional values matches quite directly how pointers work. The fact that assignment is by copy, and there is a separate move operation, is a bit of a surprise, but once explained it seems to make sense. The ease of parallel programming and the lack of problems involving undesired aliasing are seen by these ParaSail users as valuable benefits of the shift. Perhaps the bigger challenge for some is the lack of global variables in ParaSail. Eliminating global variables seems to require more restructuring than does doing without pointers. It would be possible to allow global concurrent objects in ParaSail without interfering with easy parallelization, but these would add complexity to the language and its analysis in other ways.

A second important goal was for ParaSail to enable the creation of safe, parallel programs, of significant size, with the same ease and productivity of sequential programming. For our own use, we have written two major programs in ParaSail: a compiler "back end" that reads the PSVM representation of a ParaSail program and generates the corresponding LLVM [21] instructions, and a static analyzer ("ParaScope") that does an advanced static analysis of ParaSail programs identifying all places where a precondition or postcondition might fail. These programs have confirmed the safety, flexibility, and convenience provided by pointer-free expandable objects and region-based storage management. In addition, the elimination of global variables and aliasing meant that these programs were trivial to parallelize, resulting in significant speed-ups when executed on multi-core processors because of the ability to concurrently compile or analyse the various operations that make up the typical ParaSail program. As an example, we ran the compiler back end on the 10.5k lines of code that comprise the ParaSail run time library, first using only a single core, and then second using all four virtual cores of a dual core hyper-threaded 3.5 GHz Intel i7 processor (averaged over eight runs in each configuration):





|                          | Time in seconds |       |                     |
| ------------------------ | --------------- | ----- | ------------------- |
| Number of cores          | Wall Clock      | CPU   | CPU Utilization (%) |
| Single threaded          | 179.8           | 177.2 | 98.5                |
| Hyper-threaded Dual Core | 89.5            | 306.8 | 342.9               |

This LLVM-targeted compiler back end was written by a summer intern who had not programmed in a parallel programing language before. Nevertheless, as can be seen from the table, executing this ParaSail program using multiple threads, while it did incur CPU scheduling overhead, more than made up for this overhead thanks to the parallelism "naturally" available in the program, producing a two times speed-up when going from single-threaded single core to hyper-threaded dual core. Note that "CPU" Time and Utilization in the above effectively refers to hyper-thread time and hyper-thread utilization. This simple comparison showed us, anecdotally, that a large program could be written by a parallel-programming neophyte using the inherently safe, parallel, and pointer-free constructs of ParaSail, and still achieve significant speed-up from parallel execution.

As an example of a speed up provided by implicit parallelism associated with a naive divide-and-conquer recursive Quicksort algorithm, the "functional" non-in-place recursive function Qsort, shown in section 4.1 above, was timed on sorting one million random integers, once in a single-threaded implementation, and once using parallelism inserted implicitly by the ParaSail compiler, on the same processor configuration as above and again averaged over 8 runs:

|                          | Time in seconds |       |                     |
| ------------------------ | --------------- | ----- | ------------------- |
| Number of cores          | Wall Clock      | CPU   | CPU Utilization (%) |
| Single threaded          | 77.0            | 76.0  | 98.6                |
| Hyper-threaded Dual Core | 42.6            | 142.2 | 333.8               |

Here we see that parallelism can be significant even when provided automatically in ParaSail by concurrent evaluation of complex operands of binary operations.

A third important goal was that ParaSail would raise the level of abstraction so as to increase expressivity, without incurring the kind of "abstraction bloat" that can arise in some object-oriented languages [26]. The ParaScope static analyser mentioned above was essentially a rewrite of a commercial static analyser written in Ada. As a simple comparison in expressivity, lines of code for the value propagation phase of these analysers were compared:

| Implementation Language | Source Lines of Code |
| ----------------------- | -------------------- |
| Ada                     | 31029                |
| ParaSail                | 9937                 |

The added expressivity came largely from the ease of defining and using higher-level data structures tailored to a given context, with the same economic syntax as built-in constructs, without incurring the per-abstraction-layer overhead of other





object-oriented languages. Furthermore, this significant increase in expressivity was achieved despite the lack of global variables and re-assignable pointers.

Another goal of the ParaSail design was to support real-time embedded parallel programming, with the parallel threads synchronizable via safe and efficient mutual exclusion and inter-thread signaling, while preserving bounded, predictable execution time and storage use. For these purposes, we chose to allow any module to be declared **concurrent**, as illustrated earlier with the Locked_Box, and within such a module to allow the definition of operations that provide shared read-only access, exclusive read-write access, and queued read-write access defined by a dequeue condition. This is indicated by the use of the **locked** and **queued** modifiers on parameter modes, and a **queued until** clause on a queued operation to define the dequeue condition. Together these provide implicit safe synchronization of access to shared data, along with signaling between threads, without the need for explicit lock, unlock, signal, or wait operations.

And perhaps the overarching goal of the ParaSail design was to achieve our various other goals with economy of means – to minimize the number of distinct concepts in the language without hampering expressivity. Other language designers have expressed similar goals, with the Modula-3 design team giving themselves an explicit fifty-page "complexity budget" for their reference manual [8]. Other than the reference manual size, what other measures might be used to measure inherent language complexity? In the design of ParaSail, we attempted to remove redundancy in the language design from the beginning. Many languages have adopted the notion of package or namespace to group classes or types together. In ParaSail we chose to use the same construct for grouping as we used for typing (the module). Many languages have the notion of generic templates, which are instantiated to produce non-generic units, which can then be used to create run-time entities. In ParaSail we chose to treat all modules as being parameterized, with a parameterless module being merely a special case. Each ParaSail data type is defined as an instantiation of a module, giving the actual parameters for any formal parameters of the module. And a ParaSail data object is defined as an instance of a type. There is no distinction between reference types and value types – assignment is always by value, implying having two copies of the same value when done, while a move operation is provided to move the value of one object into a second object while nulling out the value of the source object, such that there is no net increase in the amount of storage in use. Finally ParaSail has essentially one kind of operation, making no particular distinction between what other languages might call constructors, static functions, virtual functions, etc. The parameter and result profile of the operation, and whether it is declared in the interface or implementation of a module, together determine what sort of role the operation plays.

## 7 Conclusion

The original purpose of our eliminating pointers from ParaSail was the support of easy, pervasive parallelism without being restricted to the regular data-parallel computations that are common in the high-performance computing domain. From that point of view, ParaSail is a good showcase. ParaSail programs, even those manip-





ulating irregular data structures, produce a great deal of parallelism without the programmer having to make any significant effort. Almost any algorithm that is structured as a recursive walk of a tree, or as a divide and conquer algorithm such as a recursive Quicksort, will by default have its recursive calls treated as potentially parallel sub-computations. Monitoring built into the ParaSail interpreter indicates the level of parallelism achieved, and it can be substantial for algorithms not normally thought of as being embarrassingly parallel. We believe the implicit, safe, pervasive parallelism provided by ParaSail is one of its unique contributions, and this relies on the simplifications made possible by the elimination of pointers and other sources of hidden aliasing, as well as the overall economy of means, which we believe is another contribution of the ParaSail language design.

**Acknowledgements** We thank the reviewers for their helpful suggestions. Work on ParaSail has been supported by AdaCore, in part by supporting the extraordinarily productive summer internship of Justin Hendrick, who essentially single-handedly produced the initial version of the LLVM code generator. Sections 3 and 4 of this paper are expansions on information from a workshop paper that described ParaSail [35].

## About the author

**S. Tucker Taft** is VP and Director of Language Research at AdaCore, in Lexington, MA, USA. From 1990 to 1995 Tucker led the Ada 9X language design team, culminating in the February 1995 approval of Ada 95 as the first ISO standardized object-oriented programming language. Tucker is also a member of the ISO Rapporteur Group that developed Ada 2005 and Ada 2012. His specialties include programming language design, compiler implementation, high-integrity software, real-time systems, parallel programming, and model-based development. More recently Tucker has been designing and implementing the parallel programming language ParaSail, and defining parallel programming extensions for the Ada language. Contact him at taft@adacore.com.

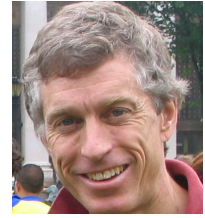